# Dynamic Dispersion Accumulation in Fiber Loops for Realizing Record-High Frequency Resolution or Ultra-Low Signal Sampling Rate in Dispersion-Based Photonics-Assisted Wideband Microwave Measurement Systems

Chi Jiang, Taixia Shi, Hang Yang, Lei Gao, Xianxin Zhang, Yiqing Liu, and Yang Chen*

Shanghai Key Laboratory of Multidimensional Information Processing, East China Normal University, Shanghai, China
E-mail: ychen@ce.ecnu.edu.cn

**ABSTRACT**
Dispersion-based photonics-assisted microwave measurement systems provide immense potential for real-time analysis of wideband and dynamic signals. However, they face two critical challenges: a difficulty in achieving high frequency resolution over a wideband analysis bandwidth, and a reliance on large-bandwidth-and-high-sampling-rate oscilloscopes to capture the resulting ultra-narrow pulses. We introduce a dynamic dispersion accumulation technique to overcome these limitations. By circulating the optical signal in fiber loops containing a dispersion-compensating fiber, we achieve a high accumulated dispersion of −215700 ps/nm. This high dispersion relaxes the required chirp rate of the chirped optical signal, enabling two distinct advantages: When the analysis bandwidth is fixed, a lower chirp rate enables a longer temporal period, yielding a record-high frequency resolution of 27.9 MHz; When the temporal period is fixed, a lower chirp rate enables a smaller bandwidth, generating a wider pulse and thus relaxing pulse sampling requirements at the expense of analysis bandwidth. This sacrifice in analysis bandwidth can be compensated by a duty-cycle-enabling technique, which holds the potential to extend the analysis bandwidth beyond 100 GHz. This work breaks the performance and hardware limitations in dispersion-based systems, paving the way for high frequency resolution, wideband microwave measurement systems that are both real-time and cost-effective.

## 1. Introduction

Fast, accurate, and wideband microwave measurement, including frequency measurement and time–frequency analysis, has become critical for advanced electronic systems that operate in increasingly crowded and dynamic spectral environments. This capability is driven by demands from key applications: In cognitive radio [1–3], fast and precise identification of idle channels across a wide spectrum is essential for maximizing spectral efficiency and enabling dynamic spectrum access; In autonomous driving [4, 5], the fast microwave measurement is essential to allocate clean frequency bands for radar and communication, thereby avoiding interference and ensuring safety; In electronic warfare [6], the fast capture and spectrum characterization of fleeting threat signals within a broad bandwidth is crucial for enabling cognitive capabilities and facilitates further intelligent decision-making, which directly enhances the antijamming ability and survivability of friendly systems. Therefore, these scenarios impose stringent requirements on the microwave measurement system, demanding a combination of large analysis bandwidth, high measurement accuracy, high frequency resolution, high temporal resolution, and fast processing speed. While conventional electronic solutions deliver excellent frequency measurement accuracy, frequency resolution, and deep signal analysis abilities through digital signal processing (DSP) [7], their performance faces a fundamental trade-off between analysis bandwidth and processing speed due to the inherent electronic bottleneck, limiting their real-time capability in wideband scenarios. Photonic solutions can overcome this critical limitation by harnessing the photonic advantages [8, 9], enabling fast microwave measurement of wideband signals.

Photonics-assisted wideband microwave measurement methods can be classified according to their mapping principles: frequency-to-power mapping (FTPM) [10–13], frequency-to-space mapping (FTSM) [14, 15], and frequency-to-time mapping (FTTM). The FTPM-based method offers a simple and efficient measurement by mapping microwave frequencies into optical or electrical signal power ratios via an amplitude comparison function [10, 11]. However, this method is limited to detecting a single-tone signal, unless it is assisted with other techniques, such as scanning receivers [12, 13]. The FTSM-based method commonly employs photonic channelization, which divides the target frequency band into sub-channels. This architecture necessitates independent processing paths for each sub-channel along with final signal synthesis, which inevitably introduces high system complexity. While this architecture achieves a large analysis bandwidth, the associated complexity complicates practical implementation [14, 15]. Given the limitations in both FTPM-based and FTSM-based methods, the FTTM-based method has attracted significant interest in recent years, primarily due to its distinct advantage in measuring multitone signals and complex signals commonly employed in practical applications. Two primary approaches have been proposed for performing FTTM: One based on optical frequency-sweeping and filtering, and the other based on dispersion. The former modulates the signal under test (SUT) onto a short-period chirped optical signal, thereby effectively imposing a time window on the SUT through the utilization of the chirped optical signal. A fixed optical filter is used to convert the chirped signals into temporal pulses. The SUT frequency is identified by detecting the temporal locations of the pulses, thus achieving FTTM [16–20]. However, this approach faces a fundamental performance trade-off among the analysis bandwidth, temporal resolution, and frequency resolution, all of which are governed by the chirp rate of the chirped optical signal and the optical filter bandwidth. The stringent requirement to match these two parameters for optimal resolution, as proposed in [21], creates an intrinsic constraint. Specifically, imposing this matched condition for optimal frequency resolution operation forces the method into one of two constrained regimes: Operating at a low chirp rate (e.g., GHz/μs to GHz/ms) enables high frequency resolution (up to MHz), but inevitably limits the analysis bandwidth or the temporal resolution. Conversely, increasing the chirp rate to the GHz/ns level improves the temporal resolution or enables wider instantaneous analysis bandwidth, but this comes at the direct expense of frequency resolution, which is constrained to the GHz level. Therefore, this approach is inherently unsuitable for application scenarios that simultaneously demand large analysis bandwidth, high temporal resolution, and high frequency resolution.

The latter approach, based on dispersion, achieves FTTM by compressing the optical signal with a quadratic phase into a narrow temporal pulse. In principle, this approach can analyze microwave signals spanning tens of GHz on a nanosecond level. Although the existing implementations typically achieve frequency resolution only at the GHz level, the dispersion-based system has the potential to achieve frequency resolution on the order of tens of MHz, thereby circumventing the inherent trade-off of the optical frequency-sweeping and filtering approach. In [22] and [23], the optical signal with a quadratic phase is obtained by stretching the pulse from a mode-locked laser using a dispersion medium. However, such implementations are constrained by the nonoverlap assumption: the stretched optical pulses must be separated by gaps to avoid temporal aliasing of SUT, inevitably leading to a loss of some SUT information. Although a short-pulse sampling technique is proposed in [24] to enable signal capture without information loss, its achievable analysis bandwidth is limited by the pulse repetition rate, reaching only 4.86 GHz. To achieve signal capture without information loss in a larger frequency range, some methods have been developed to generate the optical signal with a quadratic phase, including an optical time-lens [25, 26], fractional Talbot designs [27, 28] or directly modulating a chirped electrical signal on an optical carrier [29, 30]. They typically employ a continuous-wave (CW) laser diode, a phase modulator, and an auxiliary signal generated from an arbitrary waveform generator (AWG). In [26], a system based on an optical time lens, employing a linearly-chirped fiber Bragg grating (LCFBG) with a dispersion of 20.39 ps/nm, achieves the analysis bandwidth of 448 GHz. However, it only delivers a frequency resolution of 16 GHz. Moreover, its reliance on a sampling oscilloscope (OSC) with a bandwidth of 500 GHz for capturing the extremely narrow pulse prevents real-time analysis in practical applications. In [27], a Talbot array illumination system based on fractional Talbot designs is proposed for real-time analysis of arbitrary microwave signals. The system utilizes an LCFBG with a dispersion of 2038.49 ps/nm to achieve a frequency resolution of 2.2 GHz. However, its implementation faces constraints from demanding hardware requirements, particularly the need for a 92 GSa/s AWG to generate the Talbot array illumination signal with a $2V\pi$ amplitude and an OSC with a bandwidth of 28 GHz, which collectively limit the applicability. In [29], a V-shape chirped signal is employed, utilizing a dispersion-compensating fiber (DCF) with a dispersion of −6817 ps/nm, which achieves an analysis bandwidth of 63.2 GHz and a frequency resolution of 1.1 GHz. Furthermore, a duty-cycle-enabling technique is proposed to extend the analysis bandwidth to 252 GHz. It should be noted that the duty-cycle-enabling operation results in an unavoidable loss of SUT information. The work itself acknowledges this limitation and proposes methods in its discussion to mitigate the associated information loss. Nevertheless, further improvement of the frequency resolution is limited by the dispersion of the DCF and the 20-GHz bandwidth OSC. Analysis of these methods reveals a critical bottleneck: the frequency resolution of the dispersion-based wideband microwave measurement systems is ultimately limited by the dispersion of the dispersion medium, where a larger value is indispensable for achieving higher frequency resolution. However, the dispersion achievable from a single dispersion medium is inherently limited, which necessitates specialized approaches to deliver a high equivalent dispersion. In [30], a system utilizes a fiber loop to achieve a high equivalent dispersion of 450000 ps/nm, enabling a frequency resolution of 30 MHz. However, the analysis bandwidth is limited to only 0.53 GHz by the free spectral range (FSR) of the fiber loop and it still needs an OSC with a bandwidth of 20 GHz to capture the generated pulses. At the same time, the lack of tunability in the FSR of the fiber loop consequently limits its applicability in wideband scenarios. In summary, existing dispersion-based photonics-assisted microwave measurement systems struggle to achieve high frequency resolution while simultaneously maintaining a large analysis bandwidth. Furthermore, the generated narrow pulses necessitate the use of a large-bandwidth OSC for capturing. Thus, the dispersion-based photonics-assisted microwave measurement systems face two critical challenges: the difficulty in achieving high frequency resolution over a large analysis bandwidth, and the reliance on sampling hardware with prohibitively high sampling bandwidth and sampling rate.

In this work, we introduce a photonics-assisted wideband microwave measurement system based on dynamic dispersion accumulation, which can overcome the above two challenges. The key to performing dynamic dispersion accumulation to achieve a high accumulated dispersion is controlling the chirped optical signal passing through the DCF repeatedly in fiber loops. According to the pulse compression condition, a higher dispersion relaxes the requirement for the chirp rate of the chirped optical signal. A lower chirp rate can be realized by employing a signal with either a reduced bandwidth for a fixed temporal period or a longer temporal period for a fixed bandwidth, each offering a distinct advantage: The former broadens the compressed pulse, thereby relaxing the sampling rate requirement, while the latter enhances the frequency resolution. An experiment is performed. A DCF with a dispersion of −1685.7 ps/nm is incorporated into the fiber loop. After circulating the optical signal for 128 rounds, the accumulated dispersion reaches −215770 ps/nm, corresponding to a required chirp rate of 0.58 GHz/ns for pulse compression. In the case of using an 8-ns chirped optical signal with a bandwidth of 4.63 GHz, a compressed pulse with a pulsewidth of 179.8 ps is obtained. This enables the pulse to be captured by a 4.63-GHz OSC, thereby allowing signal acquisition at a substantially reduced sampling rate. Although the analysis bandwidth is correspondingly limited to 4.63 GHz, the duty-cycle-enabling technique can be applied, offering the potential to extend the analysis bandwidth over 100 GHz. Furthermore, when using a 64-ns chirped optical signal with a duty cycle of 1/2, a record-high frequency resolution of 27.9 MHz and an analysis bandwidth of 37.03 GHz are achieved. These results demonstrate that the dynamic dispersion accumulation technique provides a powerful solution for achieving relaxed sampling requirements or high frequency resolution. The proposed concept presents a compelling pathway to performance advancement and hardware requirements relaxation in dispersion-based photonics-assisted wideband microwave measurement systems, unlocking significant potential for future research and applications.

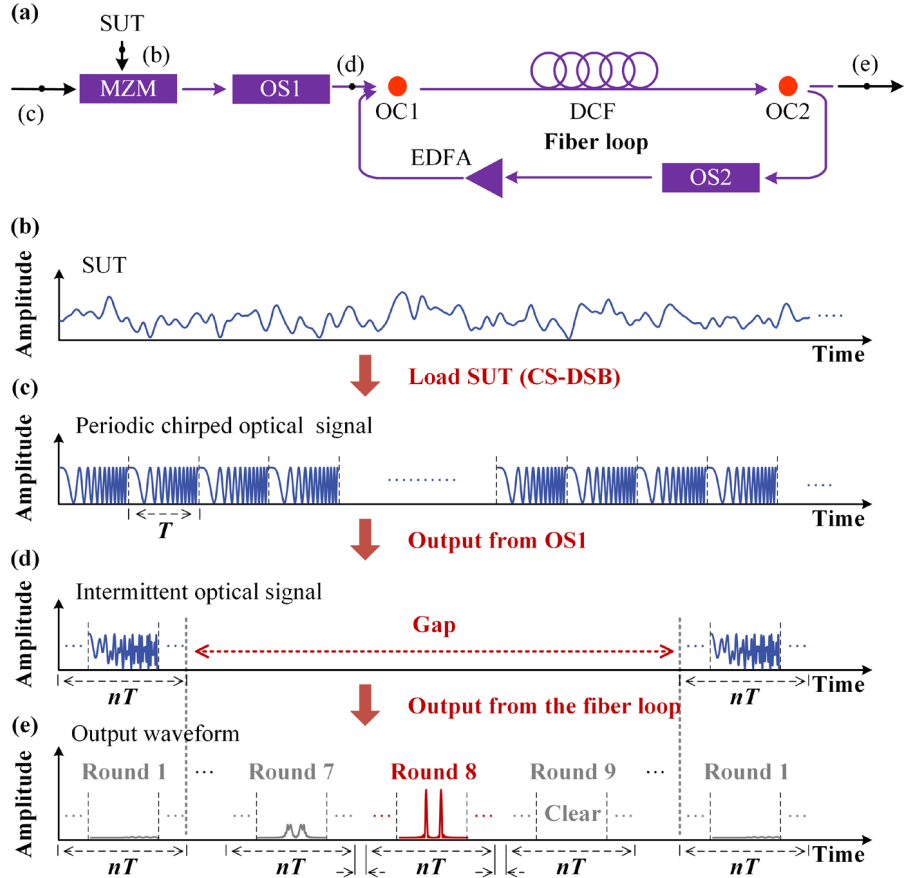

Figure 1. a) Schematic of the frequency-to-time mapping based on dynamic dispersion accumulation in fiber loops. b) Waveform of the SUT. c) Waveform of the periodic chirped optical signal. d) Waveform of the intermittent optical signal. e) Waveform of the optical signal output from the fiber loop. SUT, signal under test; MZM, Mach–Zehnder modulator; OS, optical switch; OC, optical coupler; DCF, dispersion-compensation fiber; EDFA, erbium-doped fiber amplifier.

## 2. Principle

The core of the proposed photonics-assisted wideband microwave measurement based on dynamic dispersion accumulation is to achieve a high accumulated dispersion by having a chirped optical signal circulate through a DCF within fiber loops, as illustrated in Figure 1. Figure 1a schematically illustrates the FTTM process enabled by the dynamic dispersion accumulation in fiber loops. Firstly, the SUT (Figure 1b) is loaded onto a periodic chirped optical signal (Figure 1c) at a Mach–Zehnder modulator (MZM) via a carrier-suppressed double-sideband (CS-DSB) modulation. The modulated optical signal is then gated by an optical switch (OS1) to generate an optical signal with intermittent occurrence with a temporal length of $nT$ (Figure 1d). The intermittent optical signal is injected into the fiber loop via a 50:50 optical coupler (OC1). Once injected into the fiber loop, it propagates through the DCF, where it acquires a frequency-dependent phase shift introduced by fiber dispersion. A second 50:50 coupler (OC2) then splits the optical signal output from the DCF: One path is directed to the output for detection, the other is fed back to the fiber loop for circulation. To compensate for fiber loop losses and enable stable multi-round operation, an erbium-doped fiber amplifier (EDFA) is placed after OS2 to amplify the optical signal before it re-enters OC1.

To ensure controllable operation of this circulating scheme, the two OSs must follow a coordinated timing sequence. Initially, both OS1 and OS2 are turned ON simultaneously to inject the chirped optical signal into the fiber loops. First, the ON-state duration of OS1 must be shorter than the single-round circulation delay of the fiber loop to prevent temporal overlap between successive circulations. Until the optical signal in the fiber loop is cleared, OS1 should remain in the OFF-state. Second, the ON-state duration of OS2 must be sufficiently long to allow the optical pulse to circulate for a specific number of rounds, thereby enabling precise control over the accumulated dispersion. Subsequently, OS2 is switched OFF for a duration longer than the single-round circulation delay to clear the optical signal in fiber loops, thus preventing interference with subsequently injected intermittent optical signal. Only after this clearing period can OS1 and OS2 be simultaneously turned ON-state again to initiate the next measurement cycle.

A massively high accumulated dispersion can be achieved by recirculating the intermittent optical signal for multiple rounds in fiber loops. The FTTM process completes when the chirped optical signal has circulated a predetermined number of rounds, at which point the accumulated dispersion matches the chirped optical signal's chirp rate. This results in the compression of the chirped optical signal into two symmetric temporal pulses, whose positions are linearly dependent on the input frequency, as illustrated by the red line in Figure 1e. Optical signals that undergo an unpredetermined number of circulations can not be compressed due to dispersion mismatch, as illustrated by the gray line in Figure 1e. Notably, the symmetric temporal waveform results from the CS-DSB modulation of the SUT onto the optical carrier. The analysis bandwidth could be doubled by implementing a CS-SSB modulation using a DP-MZM [29] [31].

Mathematically, the dynamic dispersion accumulation from the circulation process can be described by a linear relationship with the number of rounds. Defining the dispersion provided by a DCF as

$$D_{\text{unit}} = DL \tag{1}$$

where $D$ is the dispersion parameter of the DCF, and $L$ is the length of the DCF. The total accumulated dispersion after the optical signal undergoes $N$ circulation is given by

$$D_{\text{total}} = ND_{\text{unit}} \tag{2}$$

This linear relationship is the foundation of the dynamic dispersion accumulation. Consequently, by precisely controlling the number of circulation rounds via the two OSs, a high accumulated dispersion can be deterministically achieved. The high accumulated dispersion is the key that enables the subsequent record-resolution frequency measurement or pulse capture at a low sampling rate.

In [29] and [31], we proposed that for a chirped optical signal to be compressed into a narrow pulse to perform FTTM, its chirp rate, $k$, must satisfy the following relationship with the total dispersion:

$$k = \frac{c}{\lambda^2 D_{total}} = \frac{c}{\lambda^2 NDL} \tag{3}$$

where $c$ is the light speed in a vacuum, $\lambda$ is the center wavelength of the input optical signal. Thus, according to Equation 3, a higher total dispersion (i.e., a larger number of circulation rounds) allows for the use of a chirped optical signal with a lower chirp rate. This lower chirp rate can be achieved in two distinct configurations: when the bandwidth is fixed, it permits a longer temporal period; conversely, when the temporal period is fixed, it corresponds to a narrower bandwidth.

The significance of the former configuration is directly linked to the system's frequency resolution. As proposed in [29] and [31], we proposed that the frequency resolution of the dispersion-based microwave measurement system, $r_f$, can be expressed as:

$$r_f \approx \frac{0.886}{T} \tag{4}$$

where $T$ is the temporal period of the chirped signal. Equation 4 indicates that the frequency resolution is inversely proportional to the temporal period of the chirped optical signal. Therefore, by leveraging high accumulated dispersion to enable a low chirp rate and thus a longer temporal period, the proposed photonics-assisted wideband microwave measurement system based on dynamic dispersion accumulation would achieve a high frequency resolution. This demonstrates that the proposed approach successfully overcomes the fundamental limitation in achieving high dispersion, establishing it as a powerful new paradigm for achieving high-resolution, wideband microwave photonic measurement.

The pulsewidth of the compressed pulse can be calculated to be around

$$\tau = |0.886 / B| \tag{5}$$

where $B$ is the bandwidth of the chirped optical signal. Equation 5 indicates that the pulsewidth of the compressed pulse is inversely proportional to the bandwidth of the chirped optical signal. Therefore, by leveraging high accumulated dispersion to enable a low chirp rate and thus a smaller bandwidth, the proposed photonics-assisted wideband microwave measurement system based on dynamic dispersion accumulation would generate wider pulses, which drastically relaxes the analog bandwidth and sampling rate requirements for the real-time OSC. As a result, wideband microwave signals can be captured and analyzed using a relatively low-speed ADC, significantly reducing the system's cost and complexity.

## 3. Results

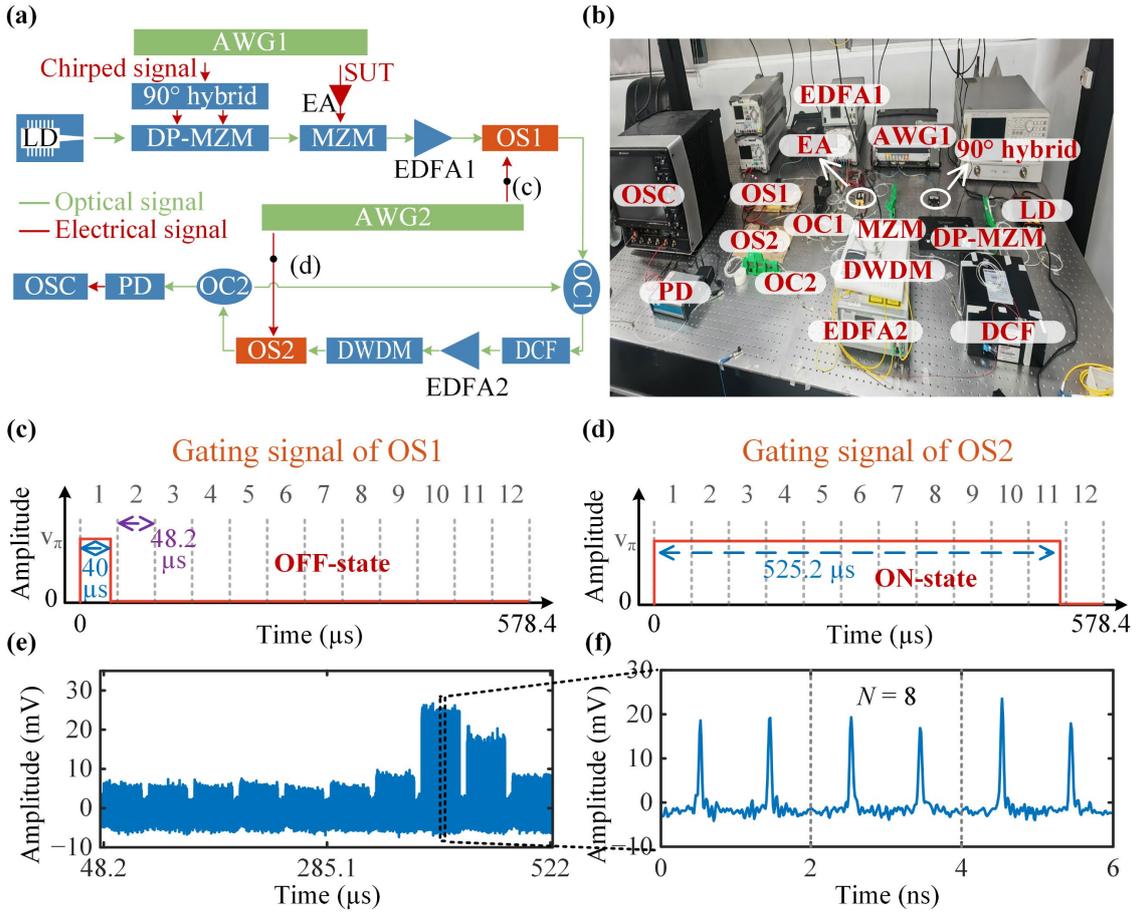

Figure 2. a) Experimental setup of the proposed photonics-assisted wideband microwave measurement system based on dynamic dispersion accumulation in fiber loops. b) Photograph of the experimental setup. Gating signal of c) OS1 and d) OS2. e) Evolution of the output waveforms from $N$ = 1 to 10. f) Three FTTM results at $N$ = 8. LD, laser diode; DP-MZM, dual-parallel Mach–Zehnder modulator; MZM, Mach–Zehnder modulator; EA, electrical amplifier; EDFA, erbium-doped fiber amplifier; OS, optical switch; AWG, arbitrary waveform generator; OC, optical coupler; DCF, dispersion-compensating fiber; DWDM, dense wavelength division multiplexing; PD, photodetector; OSC, oscilloscope.

### 3.1. Experimental Setup and Microwave Measurement Validation

The experimental setup for demonstrating the proposed photonics-assisted wideband microwave measurement system based on dynamic dispersion accumulation in fiber loops is shown in Figure 2a. Figure 2b is a photograph of the experimental setup. A CW light wave centered at 1549.64 nm from a laser diode (LD, ID Photonics CoBriteDX1-1-C-H01-FA) is employed as the optical carrier and sent to a dual-parallel Mach–Zehnder modulator (DP-MZM, Fujitsu FTM7961EX) via an optical isolator and a polarization controller. A chirped signal with a chirp rate of 9.26 GHz/ns, a frequency range from 0 to 18.52 GHz, and a period of 2 ns, generated from AWG1 (Keysight M8195A), is split into two paths with a 90° phase difference via a 90° hybrid and sent to the two radio frequency (RF) ports of the DP-MZM. In the DP-MZM, the optical carrier is carrier-suppressed single-sideband (CS-SSB) modulated to generate a chirped optical signal. According to Equation 3, to compress the chirped optical signal with a chirp rate of 9.26 GHz/ns into a narrow pulse, a total dispersion of −13485.6 ps/nm is required.

To provide this high dispersion, the chirped optical signal is subsequently injected into the fiber loop. Before injection, the chirped optical signal output from the DP-MZM is modulated by the SUT at the MZM (Fujitsu FTM7938EZ), which is based at the minimum transmission point for CS-DSB modulation. The SUT is also generated from AWG1 and is amplified by an electrical amplifier (EA, Multilink MTC5515) before applying to the MZM. The optical signal output from the MZM is amplified by EDFA1 (Amonics AEDFA-PA-35-B-FA) and then sent to OS1. Here, the OS1 is implemented by an MZM, which is biased at the minimum transmission point. The gating signal from AWG2 (Rigol DG2052) is applied to the OS1's RF port to control its state: a low-level signal maintains it at the minimum transmission point, blocking the optical signal (OFF-state), while a high-level ($V_\pi$) signal switches it to the maximum transmission point, allowing transmission of the optical signal (ON-state). The ON-state duration is set to 40 μs, as shown in Figure 2c. This duration must be shorter than the single-round fiber loop delay of 48.2 μs to prevent temporal overlap between successive circulations. Given that a single-shot FTTM requires 2 ns, this 40 μs gating window enables 20,000 individual measurements. The OFF-state duration of OS1 is set to be sufficiently long to ensure that the optical signal circulating in fiber loops is completely cleared by OS2 after reaching the predetermined rounds. Otherwise, prematurely ending the OFF-state would interfere the optical signal still circulating in the loop.

The output from OS1 is sent to a DCF with a dispersion of −1685.7 ps/nm via a 50:50 OC1. EDFA2 amplifies the optical signal from OS2 to compensate for the fiber loop losses, primarily from OS2 and the DCF. A dense wavelength-division multiplexer (DWDM) acts as an optical filter to suppress the accumulated amplified spontaneous emission (ASE) noise in the loop. The optical signal output from the DWDM is then sent to OS2, which combines with OS1 to control the number of optical signal rounds in the loop. Finally, a 50:50 OC2 split the optical signal: one path is fed back to OC1 for circulation, while the other is sent to a photodetector (PD, Coherent XPDV2120R) for optical-to-electrical conversion. The resulting electrical waveform is captured by an OSC (LeCroy WaveMaster 820Zi-B, 80 GSa/s).

To achieve the target dispersion of −13485.6 ps/nm necessary for pulse compression, the optical signal must complete eight circulation rounds in fiber loops. Furthermore, to experimentally verify that optimal compression occurs at the eighth round, the system was configured for up to ten rounds. The configuration of OS2 is illustrated as shown in the timing diagram in Figure 2d. OS2 is turned ON simultaneously with OS1 at the beginning of each measurement cycle. It remains in the ON-state to allow the optical signal to recirculate, with the output after each round being sequentially captured. Once the predetermined rounds are completed, OS2 is switched to the OFF-state. Its OFF-state duration is set to be slightly longer than the single-round circulation delay, thereby fully extinguishing the optical signal in fiber loops. Following this clearing interval, both OS1 and OS2 can be turned ON again to initiate the next measurement cycle.

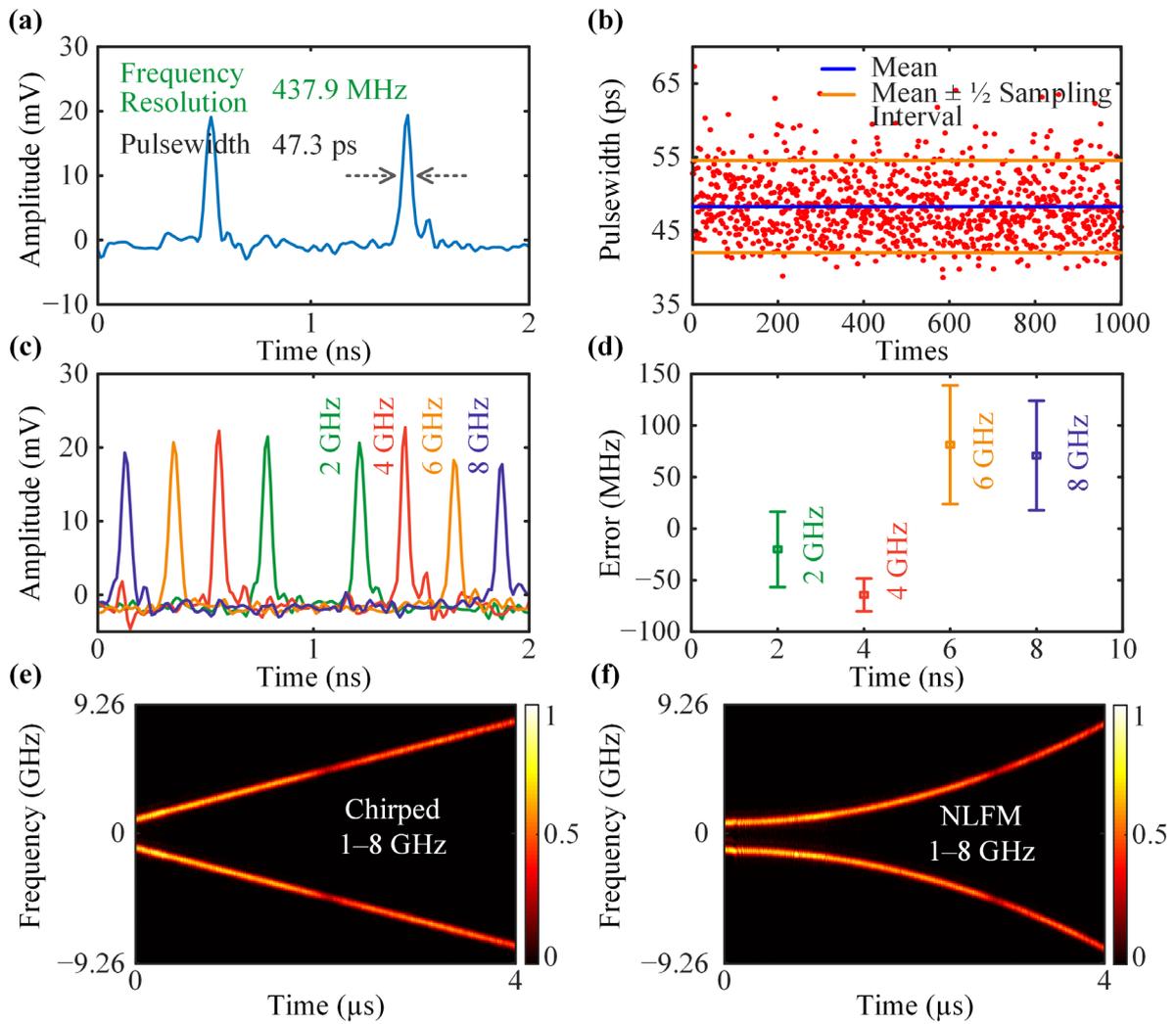

Figure 3. a) Compressed pulses in a period when a single-tone signal at 5 GHz is employed as the SUT. b) Distribution of pulsewidths over 1000 measurements. c) Compressed pulses when single-tone signals at frequencies of 2, 4, 6, and 8 GHz are employed as the SUT. d) Frequency measurement errors of the four single-tone signals. Time–frequency diagrams of e) a chirped signal, and f) an NLFM signal ranging from 1 to 8 GHz.

To validate the dynamic dispersion accumulation process, a 5 GHz single-tone signal is used as the SUT. Figure 2e shows the evolution of output waveforms after the chirped optical signal has circulated from 1 to 10 rounds in fiber loops, clearly illustrating the dynamic dispersion accumulation process. When the number of the circulation rounds is low ($N$ = 1 to 6), the accumulated dispersion remains far below the pulse compression requirement, failing to compress the chirped optical signal and resulting in low-amplitude output waveforms. Although the output optical signal amplitude increases at $N$ = 7, 9, and 10, it still cannot be compressed into a high-quality pulse. The results mean optimal pulse compression is not achieved in these cases due to dispersion mismatch. Only achieving the required dispersion at $N$ = 8, as shown in Figure 2f, the output waveform exhibits a significantly higher amplitude compared to that in other rounds, with a regular pulse shape, indicating optimal compression. These results demonstrate that the optimal pulse compression is achieved precisely at 8 rounds, in full agreement with the theoretical prediction.

With the optimal pulse compression established at $N$ = 8, we further evaluate the performance of the microwave measurement system. Figure 3a shows the compressed pulses in a period when a single-tone signal at 5 GHz is employed as the SUT. The pulsewidth is 47.3 ps, corresponding to a frequency resolution of 437.9 MHz, in close agreement with the theoretical value of 443 MHz. The stability of the pulse shape is characterized by the distribution of pulsewidths over 1000 measurements, as presented in Figure 3b. The mean pulsewidth is 48.3 ps (blue line), with most values falling within the bounds of 48.3 ± 6.25 ps, as indicated by orange lines. This fluctuation arises primarily from the 12.5 ps sampling interval of the OSC (80 GSa/s), which limits the precision of capturing narrow pulse waveforms. Figure 3c shows the captured electrical pulses in four individual measurements when single-tone signals at frequencies of 2, 4, 6, and 8 GHz are employed as the SUT, demonstrating that the pulse locations in the temporal domain are linearly dependent on the frequency of the SUT. The frequency measurement errors of 1000 measurements at the four frequencies are shown in Figure 3d. The average measurement values of the four frequencies are 1.98, 3.94, 6.08, and 8.07 GHz, respectively, while the corresponding standard deviations are 36.6, 15.9, 57.6, and 53.1 MHz, respectively. Figure 3e and 3f illustrate the time–frequency analysis results of a chirped signal with a frequency ranging from 1 to 8 GHz and a nonlinearly frequency-modulated (NLFM) signal with the same frequency range, respectively, demonstrating the capability of the proposed system for analyzing complex signals. It should be noted that a clipping process, adjusting the waveforms below 0 V to 0 V, was employed on the time–frequency results throughout the main text to reduce the influence of OSC and PD noise [29].

3.2. Wider Pulses Enabling Relaxed Sampling Requirement

As proposed in Section 2, a higher accumulated dispersion allows for the use of a chirped optical signal with a lower chirp rate. For the optical signal with a fixed temporal period, this lower chirp rate can be achieved by employing a smaller bandwidth. Consequently, according to Equation 5, the optical signal would be compressed into a wider pulse, which can relax the bandwidth and sampling rate requirements of the OSC. The following experiment is carried out to validate this advantage.

The experiment is established under a fixed chirped optical signal period of 8 ns. This period defines the system's temporal resolution and, as per Equation 4, yields a theoretical frequency resolution of 110.8 MHz. To investigate the impact of accumulated dispersion on the system performance, the total dispersion is precisely controlled by setting the number of circulation rounds in fiber loops to $N$ = 32, 64, 96, and 128. These settings produced accumulated dispersions of −53942.4, −107885, −161827, and −215770 ps/nm, respectively. According to Equation 3, these dispersions respectively correspond to required chirp rates of 2.31, 1.16, 0.77, and 0.58 GHz/ns, which can be realized using the chirped optical signal with a bandwidth of 18.52, 9.26, 6.17, and 4.63 GHz, respectively.

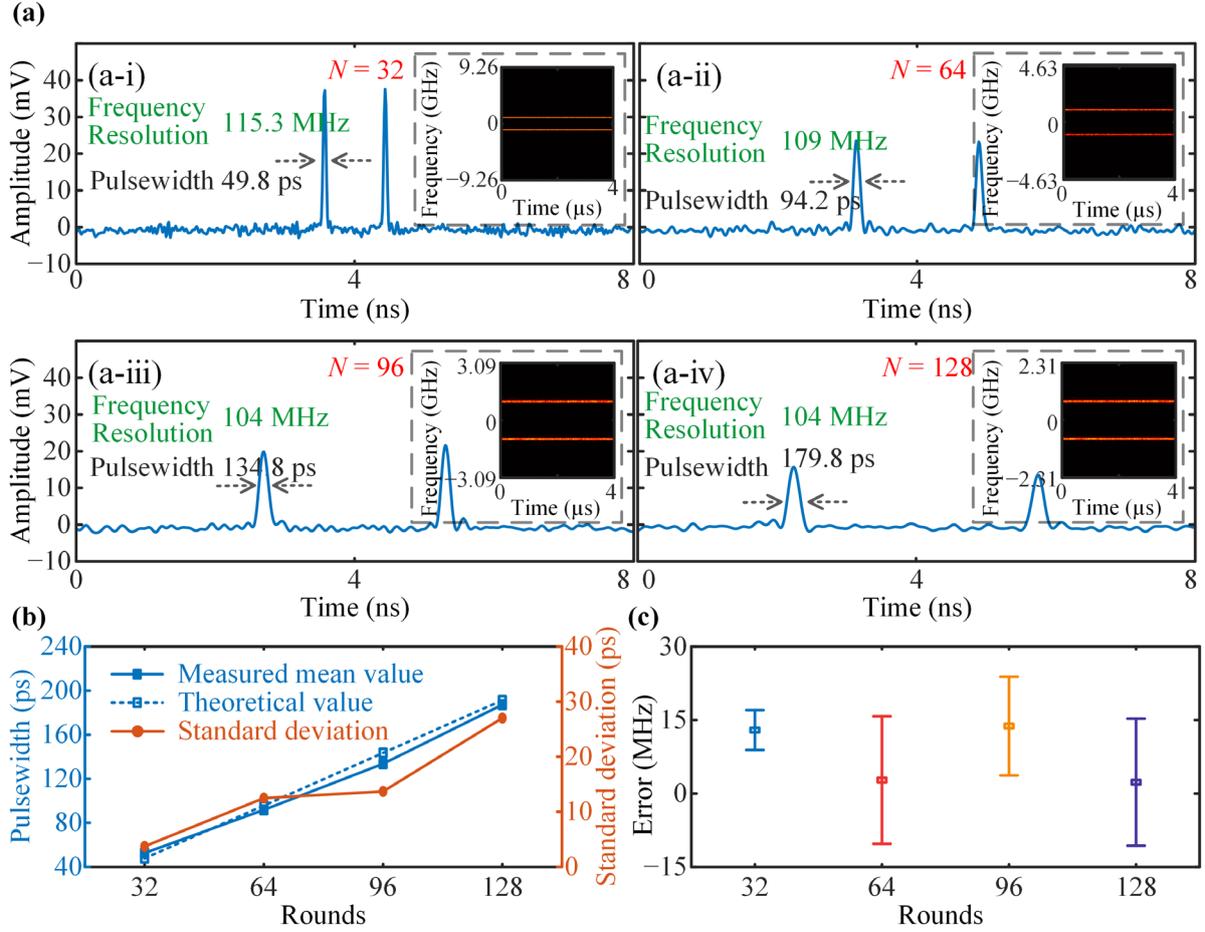

Figure 4. a) Compressed pulses in a sweep period and the time–frequency diagrams (inset) when a single-tone signal at 1 GHz is employed as the SUT. b) Statistical analysis of pulsewidths. c) Frequency measurement errors.

Firstly, it should be noted that the multiple circulations of the optical signal in fiber loops result in a low signal-to-noise ratio of the output optical signal. Thus, digital low-pass filters are applied to the waveforms captured by OSC when $N$ = 64, 96, and 128, with cutoff frequencies of 10 GHz, 7 GHz, and 5 GHz, respectively. These frequencies are set slightly above the chirped optical signal bandwidth to effectively suppress noise while preserving the signal integrity. After filtering, Figure 4a shows the compressed pulses obtained at $N$ = 32, 64, 96, and 128 in a period, respectively, when a single-tone signal at 1 GHz is employed as the SUT. The measured pulse widths are 49.8 ps, 94.2 ps, 134.8 ps, and 179.8 ps, which correspond to frequency resolutions of 115.3 MHz, 109 MHz, 104 MHz, and 104 MHz, respectively. All measured values are in excellent agreement with the theoretical resolution of 110.8 MHz, confirming the anticipated system performance. According to Equation 5, the pulsewidth of the compressed pulse is inversely proportional to the bandwidth of the chirped optical signal. Therefore, an OSC with an analog bandwidth exceeding this bandwidth is sufficient to capture the pulse without distortion. For instance, for the compressed pulse at $N$ = 128, the required OSC bandwidth is only 4.63 GHz. In contrast, without the fiber loop, the same frequency resolution is achieved only when the pulsewidth is 1.5 ps, which is presently unattainable for any real-time sampling system. This confirms that the proposed accumulated dispersion solution can produce wider pulses while maintaining its frequency resolution. The broadened pulses validate the concept for relaxing the bandwidth and sampling rate requirements of the OSC. However, this comes with a trade-off: as evidenced by the temporal location difference of the pulses in Figure 4a and the time–frequency analysis results in the insets, the analysis bandwidths reduce significantly as the circulation rounds increase from $N = 32$ to $N = 128$. This bandwidth reduction presents a practical limitation. To overcome it and achieve a larger usable analysis bandwidth, we introduce the duty-cycle-enabling technique in Section 3.3.

Given the critical role of compressed pulses in determining system performance, we statistically characterize their key parameters, including pulsewidth and temporal location, under the four dispersion conditions through 401 repeated frequency measurements. The pulsewidth governs the achievable frequency resolution, and the temporal location directly maps to the measured frequency and thus determines the frequency measurement accuracy. As summarized in Figure 4b, the measured mean pulsewidths (blue solid line) at $N = 32, 64, 96$, and 128 are 52.6, 91.8, 133.6, and 187.2 ps, respectively, which align closely with the theoretical values (blue dashed line) of 47.8, 95.7, 143.5, and 191.3 ps. The relative deviations are 10.0%, 4.1%, 6.9%, and 2.1%, respectively. This linear relationship validates the theoretical model governing pulse compression in the proposed dynamic dispersion accumulation scheme. Meanwhile, the corresponding standard deviations (orange solid line)) are 3.7, 12.5, 13.7, and 27.0 ps, representing relative standard deviations of 7.0%, 13.6%, 10.3%, and 14.4%, indicating that the pulsewidth fluctuation is sufficiently small and does not introduce a noticeable influence on frequency resolution.

Then, the temporal location of each pulse determines the measured signal frequency. The deviation of its mean value from the theoretical prediction reflects the average frequency measurement error, while its standard deviation corresponds to temporal jitter, which translates into frequency uncertainty. To visually assess the impact of pulse location on frequency accuracy, Figure 4c presents the measurement errors across different $N$ values. The average measurement frequencies are 1.0130, 1.0028, 1.0138, and 1.0023 GHz, with standard deviations of 4.1, 13.0, 10.1, and 13.0 MHz, respectively. Two key conclusions can be obtained from this data: First, the mean frequency error is consistently below 1.4%, demonstrating high accuracy across all settings; Second, the frequency uncertainty (standard deviation) remains on the order of 10 MHz despite the increase in accumulated dispersion, demonstrating stable precision.

The above results demonstrate two distinct advantages of the proposed system. First, under a fixed temporal resolution, higher accumulated dispersion allows the use of a chirped optical signal with a reduced bandwidth. This produces a wider compressed pulse, which directly relaxes the bandwidth and sampling rate requirements of the real-time OSC. Second, both the frequency resolution and measurement accuracy remain stable with increasing circulation rounds. This indicates that the accumulated noise introduced by the circulation process does not significantly degrade the pulse compression results. Consequently, the stable and predictable performance enabled by this tunable architecture is vital for practical applications.

3.3. Recover Analysis Bandwidth via a Duty-Cycle-Enabling Technique

As demonstrated in Section 3.2, a wider compressed pulse effectively relaxes the sampling rate requirements of the OSC, but at the expense of a reduced analysis bandwidth. To recover or even extend the analysis bandwidth, we adapt the duty-cycle-enabling technique from our previous work [29], which trades temporal resolution for a larger analysis bandwidth.

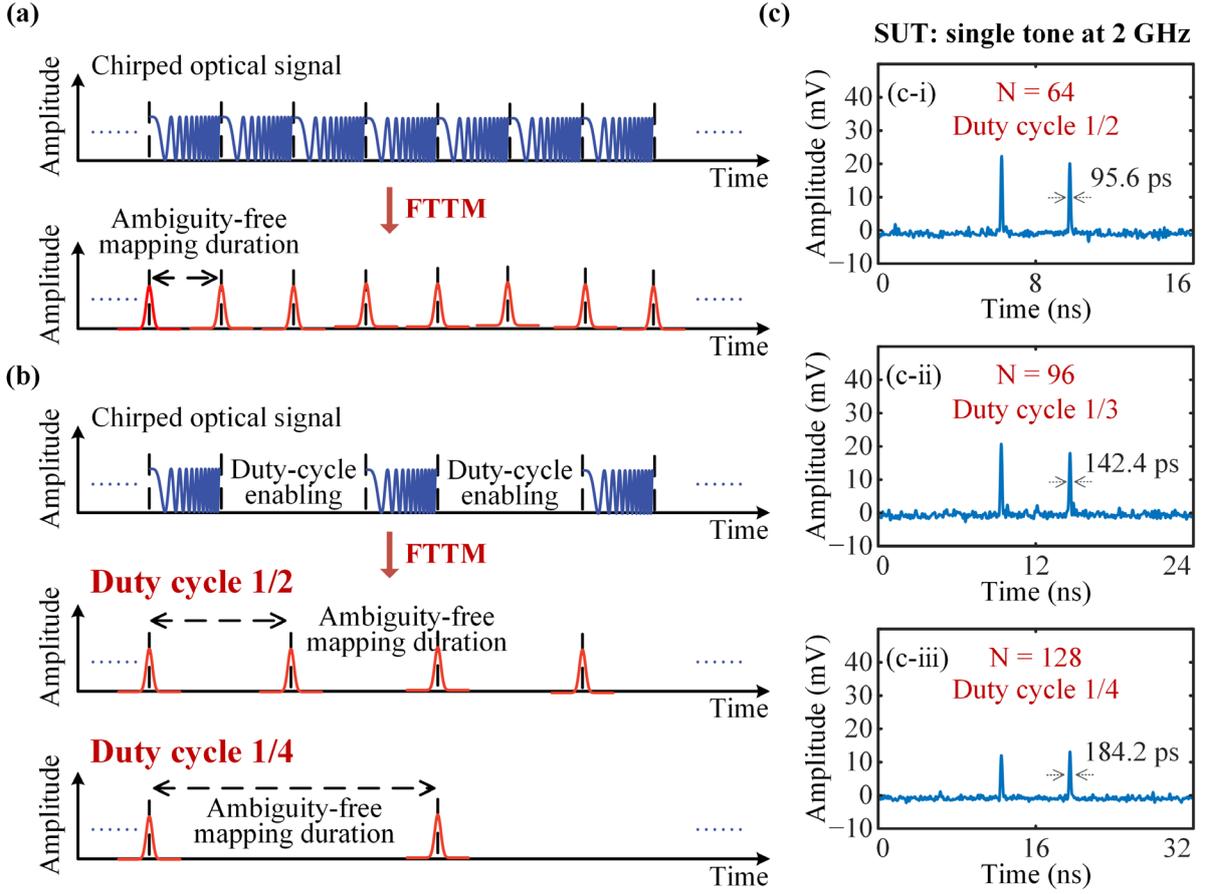

Figure 5. Schematic of the duty-cycle-enabling technique for a) a chirped optical signal without a duty cycle and b) a chirped optical signal with a duty cycle of 1/2 or 1/4. c) Attaining identical analysis bandwidth across different circulation rounds through the application of tailored duty cycles.

The principle of the duty-cycle-enabling technique is to extend the ambiguity-free mapping duration. In the case without a duty cycle, as shown in Figure 5a, the ambiguity-free mapping duration equals the period of the chirp optical signal, according to the frequency-to-time mapping principle, which defines the ambiguity-free analysis bandwidth. In Figure 5b, when a duty cycle of 1/2 or 1/4 is applied, the ambiguity-free mapping duration is effectively doubled or quadrupled, thereby increasing the ambiguity-free analysis bandwidth. This technique allows the analysis bandwidth to be flexibly controlled by simply adjusting the duty cycle of the chirped optical signal.

Figure 5c shows the application of the duty-cycle-enabling technology to optical signals compressed under different circulation rounds. The consistency proportion in compressed pulse temporal positions validates the maintenance of an identical analysis bandwidth across all configurations. The results demonstrate the capability of the duty-cycle technique to recover the analysis bandwidth sacrificed in Section 3.2. Thus, the synergy between the concepts presented in Sections 3.2 and 3.3 enables the system to maintain a large analysis bandwidth while substantially relaxing the sampling rate and bandwidth of the OSC.

The proposed microwave measurement system holds the potential for analysis bandwidths over 100 GHz, achieved by further expanding the ambiguity-free mapping duration via the duty-cycle-enabling technique. This potential is supported by [29], where a measurement bandwidth of 252.8 GHz is successfully verified.

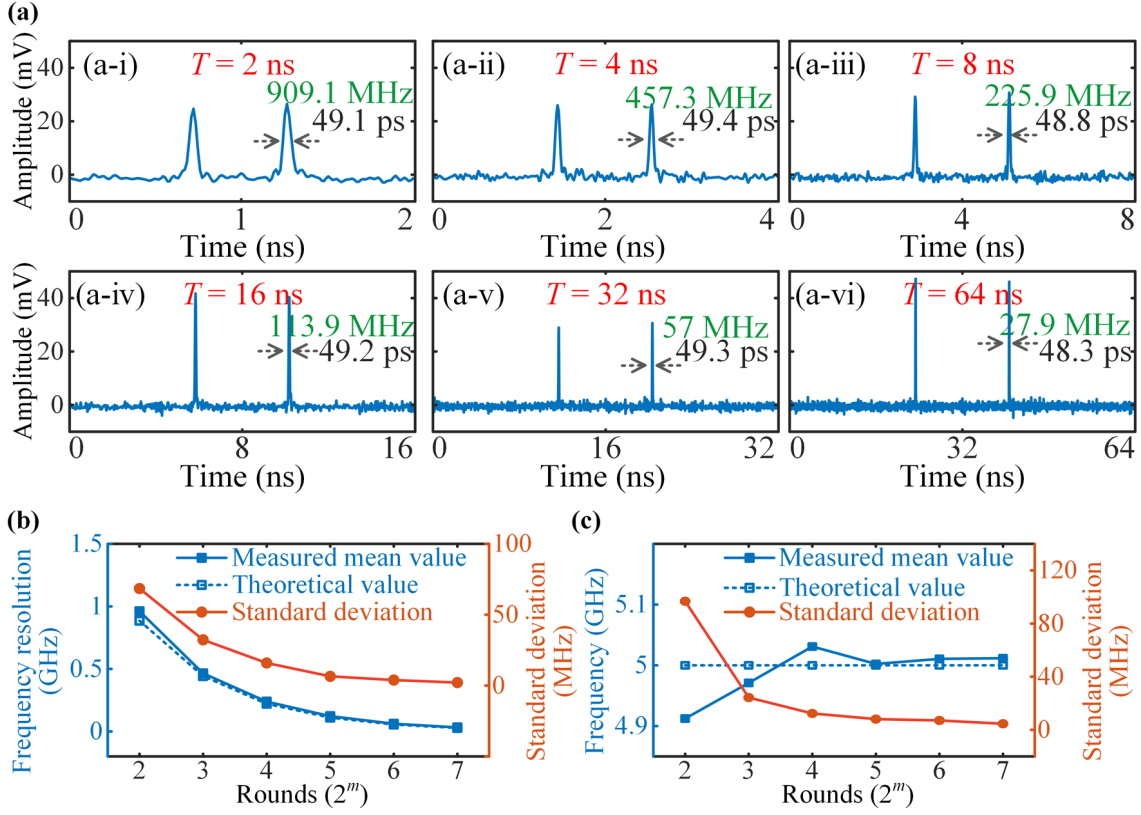

Figure 6. Experimental validation of enhanced frequency resolution. a) Compressed pulses under different circulation rounds. b) Frequency resolution and (c) frequency accuracy of 501 frequency measurements.

### 3.4. Record-High Frequency Resolution

Another approach to realizing a chirped optical signal with a lower chirp rate is to employ a longer temporal period while maintaining a fixed bandwidth. According to Equation 4, this configuration enhances the frequency resolution, and the following experiment demonstrates the achievement of a record-high frequency resolution.

We control the number of circulation rounds of the optical signal in fiber loops as $N = 2^m$ ($m$=2, 3, 4, 5, 6, 7), corresponding to accumulated dispersion of −6742.8, −13485.6, −26971.2, −53942.4, −107885, and −215770 ps/nm, respectively. For each dispersion, the required chirp rate for pulse compression is determined. To achieve a consistent analysis bandwidth of 37.03 GHz by applying a duty cycle of 1/2, the periods of the chirped optical signal should be set to $T$ = 2, 4, 8, 16, 32, and 64 ns, respectively.

Figure 6a shows the compressed pulses in a period when a single-tone signal at 5 GHz is employed as the SUT under different temporal lengths of the chirped optical signal. The consistency proportion in their temporal positions validates the maintenance of an identical analysis bandwidth across all configurations. Moreover, due to the analysis bandwidth being identical, a longer period means that a given time interval corresponds to a smaller frequency span. Therefore, despite the similar pulsewidths observed, the frequency resolution is significantly improved with the increase of the signal period. The record-high frequency resolution of 27.9 MHz is obtained when $T$ = 64 ns, as shown in Figure 6a-vi, which coincides with the theoretical value of 27.7 MHz calculated by Equation 4. In contrast, without the fiber loop and under the same analysis bandwidth, while a temporal resolution of 0.25 ns can be achieved, the corresponding frequency resolution remains as high as 3.54 GHz. This result confirms a dramatic improvement in frequency resolution accomplished by this work.

**Table 1. Comparison of different FTTM-based photonics-assisted microwave measurement systems**

| | Method | Key technique | Analysis bandwidth | Frequency resolution | Temporal resolution | Absolute dispersion @ wavelength | OSC bandwidth |
|---|---|---|---|---|---|---|---|
| Ref. [16] | frequency-sweeping and filtering | SBS effect | 12 GHz | 60 MHz | 2 μs | no need | 3 GHz |
| Ref. [18] | | Transient SBS effect | 20 GHz | 99 MHz | 100 ns | | no mentioned |
| Ref. [26] | Dispersion | optical time lenses | 448 GHz | 16 GHz | 62.5 ps | 20.39 ps/nm @ 1550 nm | 500 GHz |
| Ref. [27] | | Talbot array illumination | 92 GHz | 2.2 GHz | 1.5 ns | 2038.49 ps/nm @ 1550 nm | 28 GHz |
| Ref. [29] | | V-shape chirped signal and duty-cycle-enabling | 252 GHz | 1.1 GHz | 13.75 ns | 6817 ps/nm @ 1546.6 nm | 20 GHz |
| Ref. [30] | | optical time lenses and fiber loop | 0.53 GHz | 30 MHz | 30 ns | 450000 ps/nm @ - | 20 GHz |
| This work | | dynamic dispersion accumulation | 4.63 GHz | 104 MHz | 8 ns | 215700 ps/nm @ 1549.64 nm | 4.63 GHz (theoretical) |
| | | | 37.03 GHz | 27.9 MHz | 64 ns | | 20 GHz |

To further evaluate the frequency measurement performance under different signal periods, we present a statistical analysis of 501 frequency measurements. In Figure 6b, the blue dashed line is the theoretical frequency resolution, the blue solid line is the mean of the measured frequency resolution, and the orange solid line is the standard deviation of the frequency resolution. It can be observed that more circulation rounds lead to a higher frequency resolution, and the reduced standard deviation indicates that the frequency resolution also becomes more stable. In Figure 6c, the blue dashed line is the theoretical frequency, the blue solid line is the mean of the measured frequency, and the orange solid line is the standard deviation of the frequency measurement. The results demonstrate that the frequency measurement grows more accurate and stable as the number of circulation rounds increases. This improvement is a direct consequence of the increased accumulated dispersion, which necessitates a lower chirp rate and thereby a longer temporal period, $T$, for a fixed bandwidth. A longer $T$ means the same analysis bandwidth is mapped onto a longer temporal interval, leading to a fixed absolute pulsewidth corresponding to a smaller frequency range (a higher frequency resolution). Therefore, both the frequency resolution and the measurement accuracy are enhanced together.

## 4. Discussion

### 4.1. Comparison With Existing Methods

To highlight the advantages of the proposed photonics-assisted wideband microwave measurement system based on dynamic dispersion accumulation in fiber loops, Table 1 compares the key performance of this work with other representative FTTM-based systems previously reported. The comparison reveals the following key outcomes:

1) SBS-based systems, as demonstrated in [16] and [18], typically achieve excellent frequency resolution at tens of MHz, but exhibit poor temporal resolution on the hundreds of nanoseconds to microseconds level, which restricts their applicability in analyzing rapidly varying non-stationary signals. In comparison, most dispersion-based systems excel in temporal resolution (ns level) and can offer immense analysis bandwidths exceeding hundreds of GHz[26][27][29][30]. However, their frequency resolution is confined to the GHz level, constrained by the limited dispersion available from a single dispersion medium. Although the system in [30] utilizes a fiber loop to deliver a large dispersion and achieves a frequency resolution of 30 MHz, its analysis bandwidth is fundamentally restricted by the free spectral range of the fiber loop and is difficult to further extend. In this work, by leveraging the dynamic dispersion accumulation, we achieve an exceptionally high accumulated dispersion of −215700 ps/nm, leading to a record-high frequency resolution of 27.9 MHz among dispersion-based microwave measurement systems. Moreover, the analysis bandwidth can reach up to 37.03 GHz and can be flexibly adjusted using the duty-cycle-enabling technique, offering the potential for analysis bandwidths over 100 GHz.

2) The high accumulated dispersion can relax the optical pulse sampling requirement. By producing significantly broadened compressed pulses, the proposed system permits signal sampling with an OSC bandwidth of only 4.63 GHz, which is markedly lower than the 20–500 GHz typically needed in other dispersion-based systems[26][27][29][30]. This advantage opens up opportunities for constructing high-performance measurement systems with significantly reduced cost.

4.2. Considerations in practical application

The results demonstrated in Section 3 provide an experimental validation for the concept of photonics-assisted wideband microwave measurement system based on dynamic dispersion accumulation. However, this method also introduces a limitation: the accumulation propagation delay from each circulation round results in a system dead time, which precludes continuous signal analysis. As illustrated in Figure 1b, a temporal gap must be reserved to allow each injected signal segment to circulate without interference. The more the number of circulation rounds required, the greater the accumulated delay and the associated dead time. In practical applications, extended dead time directly compromises the real-time analysis capability.

To mitigate this limitation and enhance the practicality of the proposed system, we propose two potential solutions. First, replacing the DCF with an LCFBG would drastically reduce the loop delay. In this work, the DCF introduces a delay of 42.8 μs per round, whereas an LCFBG could reduce this to the nanosecond scale, thereby significantly shortening the dead time for the same number of circulations. Second, employing a dispersion medium with a higher dispersion per unit length would reduce the number of circulation rounds required to achieve a target total dispersion, consequently lowering the overall dead time. Both solutions offer viable pathways to improve system responsiveness.

Finally, the dynamic dispersion accumulation method introduced in this work establishes a highly flexible and scalable system for photonic-assisted wideband microwave measurement. By circulating the optical signal in fiber loops, we achieve a large dispersion that far exceeds any single dispersion medium. This accumulation-based concept provides a flexible method to control key system performance, enabling either record-high frequency resolution or ultra-low sampling rate application, as experimentally verified. The dynamic dispersion accumulation technique fundamentally frees system design from fixed dispersion limitations, offering a versatile pathway toward next-generation microwave measurement systems.

5. Conclusion

In this work, we introduced and experimentally validated a photonics-assisted wideband microwave measurement system based on dynamic dispersion accumulation, which effectively addresses the two critical challenges in conventional dispersion-based photonics-assisted microwave measurement systems: the difficulty in achieving high frequency resolution over a large analysis bandwidth, and the reliance on sampling hardware with an extremely high sampling rate. By circulating the optical signal in fiber loops, we achieved a high accumulated dispersion of −215770 ps/nm. This high dispersion enables the system to operate with two key advantages: delivering a record-high frequency resolution of 27.9 MHz using a chirped optical signal with a period of 64 ns (duty cycle of 1/2) and a bandwidth of 18.52 GHz; or generating a significantly wider pulsewidth of 179.8 ps using a chirped optical signal with a temporal length of 8 ns and a bandwidth of 4.63 GHz. The wider pulse relaxes the bandwidth requirement of the OSC to just 4.63 GHz, thereby enabling ultra-low sampling rate. Although the analysis bandwidth is limited to 4.63 GHz, applying a duty-cycle-enabling technique recovers the effective analysis bandwidth to 18.52 GHz. Critically, this work proposed a flexible design idea, allowing the system to be optimized for resolution, speed, or bandwidth as needed in dispersion-based microwave measurement. The fiber loop enables high and tunable dispersion, freeing frequency resolution and sampling rate from fixed dispersion medium limits. The duty-cycle technique then recovers any bandwidth sacrificed for lower sampling rates. This work establishes a practical and high-performance framework for wideband microwave frequency measurement and time–frequency analysis, paving the way toward real-time, high-resolution, and cost-effective systems for future wideband applications.

**Acknowledgements**
National Natural Science Foundation of China under Grant 62371191 and Grant 62401207, Shanghai Oriental Talent Program under Grant QNJY2024007, Shanghai Aerospace Science and Technology Innovation Fund under Grant SAST2022-074, Science and Technology Commission of Shanghai Municipality under Grant 22DZ2229004.